\documentclass[twocolumn]{autart}    

\usepackage{graphicx}          

\usepackage{amssymb}
\usepackage{amsmath}
\usepackage{array}
\usepackage{multirow}
\usepackage{siunitx}
\usepackage{booktabs}
\usepackage{graphicx}
\usepackage{cite}
\usepackage{bm}  
\usepackage{color}   
\usepackage{soul}    
\usepackage{natbib}                      
\allowdisplaybreaks[4]

\begin{document}
	
	\begin{frontmatter}
		
		\title{Distributed Nash Equilibrium Seeking with \\
			 Stochastic Event-Triggered Mechanism} 
		
		
		\author[HKUST]{Wei Huo}\ead{whuoaa@connect.ust.hk},    
		\author[ETH]{Kam Fai Elvis Tsang}\ead{kfetsang@kth.se},               
		\author[HKUST]{Yamin Yan \thanksref{footnoteinfo}}\ead{eeymyan@ust.hk},  
		\author[ETH]{Karl Henrik Johansson}\ead{kallej@kth.se},
		\author[HKUST]{Ling Shi}\ead{eesling@ust.hk}
		
		\address[HKUST]{Department of Electronic and Computer Engineering, Hong Kong University of Science and Technology, Hong Kong}  
		\address[ETH]{Division of Decision and
			Control Systems, KTH Royal Institute of Technology, Stockholm, Sweden}             
		\thanks[footnoteinfo]{Corresponding author.}          
		
		\begin{abstract}                          
			In this paper, we study the problem of consensus-based distributed Nash equilibrium (NE) seeking where a network of players, abstracted as a directed graph, aim to minimize their own local cost functions non-cooperatively. Considering the limited energy of players and constrained bandwidths, we propose a stochastic event-triggered algorithm by triggering each player with a probability depending on certain events, which improves communication efficiency by avoiding continuous communication. We show that the distributed algorithm with the developed event-triggered communication scheme converges to the exact NE exponentially if the underlying communication graph is strongly connected. Moreover, we prove that our proposed event-triggered algorithm is free of Zeno behavior. Finally, numerical simulations for a spectrum access game are provided to illustrate the effectiveness of the proposed mechanism by comparing it with some existing event-triggered methods.
		\end{abstract}
		
		
	\end{frontmatter}
	
	\section{Introduction} \label{sec: introduction}
	The prevalence of applications of game theory varies from power grids~(\cite{wang2021distributed}), mobile ad-hoc networks~(\cite{stankovic2011distributed}), resource allocation~(\cite{rahman2019efficient}) and social networks~(\cite{ghaderi2014opinion}), etc., capturing competition characteristics among different parts. In non-cooperative games, each self-interest player intends to maximize or minimize its local objective function which is often in conflict with other players. A Nash equilibrium (NE) in such games presents a rigorous mathematical characterization of desirable and stable solutions to the games and has attracted a considerable amount of interest in past decades. 
	
	With the rapid development of large-scale networks, traditional centralized frameworks for NE seeking algorithms where all players access all opponents' actions suffer from limited scalability and substantial computation cost~(\cite{govindan2003global, frihauf2011nash, kannan2012distributed}). 
	In view of this, distributed NE seeking in a non-cooperative game where players only communicate with their neighbors has shown theoretical significance and practical relevance in recent years. In discrete-time settings,~\cite{salehisadaghiani2016distributed} developed an asynchronous gossip-based method for seeking a NE with almost sure convergence, but diminishing step sizes slowed down the convergence. Later,~\cite{salehisadaghiani2019distributed} utilized an alternating direction method of multipliers approach to achieve the NE with constant step sizes. 
	For continuous-time cases,~\cite{gadjov2018passivity} presented a passivity-based algorithm to obtain the NE over networks by leveraging incremental passivity properties of the pseudo-gradient.~\cite{ye2017distributed} proposed a consensus-based approach to seek the NE exponentially. 
	
	The above-mentioned conventional distributed NE seeking algorithms require continuous communication, causing a high communication burden. Therefore, these algorithms can be impractical in physical applications. Especially for some embedding networks equipped with energy harvesting, the energy of each player can be a scarce resource and needs to be closely monitored and controlled. 
		A motivating application is the spectrum access game in energy-harvesting body sensor networks (BSNs).
		Multiple BSNs compete to share the bandwidth in a cognitive radio network, and they use the allocated spectrum to transmit physiological data to a remote healthcare center~(\cite{niyato2007game}). Each selfish BSN intends to minimize its own transmission cost and receive the best health service by choosing appropriate spectrum size. To achieve the NE distributively, BSNs need to interact with their neighbors to compensate for the lack of global information on others' strategies. However, continuous communication excessively consumes scarce energy harvested from the ambient environment. Thus, there is a need for novel communication-efficient algorithms seeking the NE to save the harvested energy.

	Event-triggered mechanism has gained popularity throughout the control community since it can reduce the communication burden by filtering out unnecessary information transmission~(\cite{wu2012event}). 
	As for non-cooperative games,~\cite{shi2019distributed} proposed an edge-based event-triggering law in discrete-time aggregative game. However, the convergence speed is slow owing to the diminishing step size.\textcolor{blue}{~\cite{yu2022distributed} designed a static event-triggering law with a decaying threshold, and recently,~\cite{xu2022hybrid} proposed a fully distributed edge-based adaptive dynamic event-triggered scheme for undirected networks. Nonetheless, algorithms in~\cite{yu2022distributed, xu2022hybrid} only converge to the neighborhood of the NE instead of the exact NE.}\textcolor{blue}{~\cite{liu2023predefined} constructed an adaptive event trigger and a time base generator to achieve predefined-time convergence with an arbitrarily small error.}~\cite{zhang2021distributed} have successfully applied the dynamic event-triggered method from~\cite{yi2018dynamic} to distributed games and proven that the algorithm converges to the exact NE. 
	However, all the existing works focus on deterministic event-triggered algorithms that precisely specify the triggering times for each player. 
	
	Recently,~\cite{tsang2019zeno, tsang2020distributed} extended deterministic event triggers to stochastic versions by defining the triggering time more loosely and achieved a better trade-off between communication effort and convergence performance in multi-agent consensus and decentralized unconstrained optimization in undirected networks.
	Nonetheless, the existing stochastic event-triggering laws cannot be directly applied to distributed NE seeking problems since the cost function of each player in a non-cooperative game is coupled with the actions of other players. Due to the complex information exchange setting in distributed NE seeking, the design of the stochastic event-triggered mechanism and the convergence analysis encounter more difficulties. 
		Constrained action sets should also be considered. To the best of the authors' knowledge, there is no stochastic event-triggered mechanism designed for distributed constrained NE seeking problems.
	
	All the above motivates us to develop a stochastic event-triggered algorithm for a multi-agent system to seek the NE in a distributed constrained game. The main contributions of this paper are summarized in the following:
	\begin{itemize}
		\item [1)] We propose a novel stochastic event-triggered distributed NE seeking algorithm for constrained non-cooperative games in directed networks.
		
		\item [2)] We show that the developed algorithm converges to the exact NE exponentially. Furthermore, we prove that the algorithm is free of Zeno behavior, validating its feasibility.
		\item [3)] 
		Simulation results for the spectrum access game in BSNs demonstrate the advantage of the proposed algorithm in better balancing the communication consumption and convergence properties than deterministic ones.
		
	\end{itemize}
	
	The remainder of this paper is organized as follows. In Section~\ref{sec: preliminaries}, some preliminaries are provided. Then the problem formulation about the distributed NE seeking under an event-triggered mechanism is presented in Section~\ref{sec: formulation}. In Section~\ref{sec: results}, a stochastic event-triggered algorithm is proposed first, and then the convergence, together with a guarantee on the exclusion of Zeno behavior are analyzed. Simulations are given in Section~\ref{sec: simulation} to illustrate the effectiveness of the proposed algorithm. Finally, conclusions are offered in Section~\ref{sec: conclusions}.  
	
	\emph{Notations}:
	In this paper, $\mathbb{R}$ and $\mathbb{R}^{N}$ represents the set of real numbers and $N$-dimensional real vector, respectively. $X \succ 0$ means the matrix $X$ is positive definite. 
	$\text{diag} \{a_{1}, a_{2}, \dots, a_{N}\}$ denotes an $N \times N$ diagonal matrix with elements $a_{1}, a_{2}, \dots, a_{N}$.
	The matrix $I_{N} \in \mathbb{R}^{N \times N}$ represents the identity matrix and $\mathbf{1}_{N} \in \mathbb{R}^{N}$ denotes a vector with all elements being $1$. The operator $\left \| \cdot \right \|$ is the induced $2$-norm for matrices and the Euclidean norm for vectors. 
	For any vector $\mathbf{v} \in \mathbb{R}^{N}$, $\mathbf{v}^{T}$ represents its transpose. $P(E)$ means the probability of the event $E$ happening.
	For any two matrices, $A \in \mathbb{R}^{n \times m}$, $B\in \mathbb{R}^{p \times q}$, $A \otimes B \in \mathbb{R}^{np \times mq}$ is the Kronecker product of $A$ by $B$.
	
	\section{Preliminaries} \label{sec: preliminaries}
	\subsection{Game theory}
	\begin{defn}
		A game is defined as a tuple $\boldsymbol{\Gamma} = \{ \mathcal{P}, \mathcal{X}, f \}$, where $\mathcal{P}=\{1, 2, \dots, N\}$ is the set of players, $\mathcal{X} = \mathcal{X}_{1} \times \mathcal{X}_{2} \times \cdots \times \mathcal{X}_{N}$, $\mathcal{X}_{i}\subseteq \mathbb{R}$ is the action set of the $i$th player, and $f = \{f_1, f_2, \dots, f_N\}$, $f_{i}: \mathbb{R}^{N} \to \mathbb{R}$ is the cost function of the player~$i$.
	\end{defn}
	
	\begin{defn}
		An NE is defined as an action profile $\mathbf{x}^{*} = [x_1^{*}, x_2^{*}, \dots, x_N^{*}]^{T} \in \mathcal{X}$ if $f_{i}(x_{i}^{*}, \mathbf{x}_{-i}^{*}) \leq f_{i}(x_{i}, \mathbf{x}_{-i}^{*})$, ${\forall} i \in \mathcal{V}$, where $x_{i} \in \mathcal{X}_{i}$ and $\mathbf{x}^{*}_{-i} = [x_1^{*}, x_2^{*}, \dots, x_{i-1}^{*}, x_{i+1}^{*}, \dots, x_{N}^{*}]^{T}$.
	\end{defn}
	
	\subsection{Graph theory}
		For a directed graph defined as $\mathcal{G} = \left (\mathcal{V}, \mathcal{E}\right )$, $\mathcal{V} = \{1, 2, \dots, N\}$ is the set of nodes, and $\mathcal{E}$ represents the set of edges. 
		Each edge $\left (i,j\right ) \in \mathcal{E}$ describes an available communication link from player~$j$ to player~$i$.
		The adjacency matrix $A=[a_{ij}] \in \mathbb{R}^{N \times N}$ indicates the underlying topology of $\mathcal{G}$, where $a_{ij}>0$ if $\left (i,j\right ) \in \mathcal{E}$, and $a_{ij} = 0$, if $\left (i,j\right ) \notin \mathcal{E}$. 
	
	
		The degree matrix is defined as $D=\text{\normalfont diag} \{d_{1}^{\text{in}}, d_{2}^{\text{in}}, \dots, d_{N}^{\text{in}}\}$, where $d_{i}^{\text{in}} = \sum_{j=1}^{N} a_{ij}$. The Laplacian matrix is then defined as $L=D-A$. $\mathcal{G}$ is said to be strongly connected if, for any node, there exists a directed path to every other node.  
	
	The following lemma about strongly connected directed graphs is essential for our analysis~(\cite{zhang2021distributed}):
	\begin{lem}
		$(L \otimes I_{N} + B_{0})$ is a non-singular $M$-matrix if and only if $\mathcal{G}$ is a directed and strongly connected graph, 
		where $B_{0} = \text{\normalfont diag}\{a_{11}, \dots, a_{1N}, a_{21}, \dots, a_{2N}, \dots, a_{N1}, \dots, \\a_{NN}\}$.  
There exist positive definite matrices $P,Q \succ 0$ such that
		\begin{equation} \label{PQ_lemma}
			\left (L \otimes I_{N} + B_{0}\right )^{T} P + P\left (L \otimes I_{N} + B_{0}\right ) = Q.
		\end{equation}
	\end{lem}
	
	\subsection{Projection operator}
		A set $\mathcal{X} \subseteq \mathbb{R}^{N}$ is convex if $c\mathbf{v}_{1} + (1-c)\mathbf{v}_{2} \in \mathcal{X}$, for any $\mathbf{v}_{1}, \mathbf{v}_{2} \in \mathcal{X}$ and any $c \in [0,1]$. For a closed convex set $\mathcal{X}$, the projection operator $\mathbb{P}_{\mathcal{X}}(\cdot): \mathbb{R}^{N} \to \mathcal{X}$ is defined as $\mathbb{P}_{\mathcal{X}}(\mathbf{v}) = \mathop{\arg\min}\limits_{\mathbf{z} \in \mathcal{X}} \left\| \mathbf{v}-\mathbf{z} \right\|$.

	\begin{lem}(\cite{facchinei2003finite}) For a closed convex set $\mathcal{X} \subseteq \mathbb{R}^{N}$, the projector $\mathbb{P}_{\mathcal{X}}(\cdot)$ is non-expansive, i.e., for any $\mathbf{v}_{1}, \mathbf{v}_{2} \in \mathcal{X}$,
		\begin{equation*}
			\left\| \mathbb{P}_{\mathcal{X}}(\mathbf{v}_{1}) - \mathbb{P}_{\mathcal{X}}(\mathbf{v}_{2})  \right\| \leq 
			\left\| \mathbf{v}_{1} - \mathbf{v}_{2} \right\|.
		\end{equation*}
	\end{lem}
	\section{Problem Formulation} \label{sec: formulation}
	Consider a non-cooperative multi-agent system with $N>1$ players represented by a strongly connected directed graph $\mathcal{G} = \left (\mathcal{V}, \mathcal{E}\right )$, where each selfish player~$i$ intends to minimize its own cost function, 
	\begin{equation} \label{eq: DG}
		\min_{x_{i} \in \mathcal{X}_{i}} f_{i}(x_{i}, \mathbf{x}_{-i}),
	\end{equation}
	where $\mathcal{X}_{i}$ is a closed convex set, and $f_{i}$ is the convex cost function of player~$i$ satisfying the following assumptions:
	\begin{assum} \label{existence}
		$f_{i}(\mathbf{x})$ is twice continuously differentiable and 
		$\frac{\partial f_{i}}{\partial x_{i}}(\mathbf{x})$
		is globally Lipschitz for all $i \in \mathcal{V}$, that is, there exists a constant $l_{i}>0$ such that $\left \| \frac{\partial f_{i}}{\partial x_{i}}(\mathbf{x}) - \frac{\partial f_{i} }{\partial x_{i}}(\mathbf{y})\right \| \leq l_{i} \left \| \mathbf{x} - \mathbf{y}\right \|$.
	\end{assum}
	
	\begin{assum} \label{unique}
		There exists a constant $\mu>0$ such that $(\mathbf{x} - \mathbf{y} )^{T} (F(\mathbf{x}) - F(\mathbf{y})) \geq \mu \left \| \mathbf{x} - \mathbf{y} \right \|^{2}$ for $\mathbf{x}, \mathbf{y} \in \mathbb{R}^{N}$, where 
		$F(\mathbf{x}) = \left[\frac{\partial f_{1}}{\partial x_{1}}(\mathbf{x}), \frac{\partial f_{2}}{\partial x_{2}}(\mathbf{x}), \dots,\frac{\partial f_{N}}{\partial x_{N}}(\mathbf{x}) \right]^{T} \in \mathbb{R}^{N}$ denotes the pseudo-gradient (the stacked vector of all players' partial gradients w.r.t. local cost functions).
	\end{assum}
\begin{rem}
		Under Assumption~\ref{existence}, the NE of the game~\eqref{eq: DG}, $\mathbf{x}^{*}$, is equivalent to the solutions to the varational inequality $VI(\mathcal{X}, F)$ which satisfies $(\mathbf{x} - \mathbf{x}^{*})^{T}F(\mathbf{x}^{*}) > 0, \forall \mathbf{x} \in \mathcal{X}$~(\cite{facchinei2003finite}). Assumption~\ref{unique} implies that $VI(\mathcal{X}, F)$ has at most one solution~(\cite{facchinei2007generalized}). 
		Thus, the existence and uniqueness of the NE of~\eqref{eq: DG} follows and $\mathbf{x}^{*}$ satisfies
		\begin{equation} \label{eq: NE_property}
			\mathbf{x}^{*} = \mathbb{P}_{\mathcal{X}}(\mathbf{x}^{*}- \textcolor{blue}{\tilde{\alpha}} F(\mathbf{x}^{*})), \ \forall \textcolor{blue}{\tilde{\alpha}} > 0.
		\end{equation}
	\end{rem}
	In a distributed setting, each player communicates with its neighbors to obtain partial information on the others' actions. We consider a leader-follower-based consensus control algorithm with projected gradient play dynamics~(\cite{ye2017distributed, liang2022exponentially}):
	\begin{align}
		\label{x_dyn} \dot{x}_{i} (t) =&  \mathbb{P}_{\mathcal{X}_{i}} \left( x_{i}(t) - \alpha \frac{\partial f_{i}}{\partial x_{i}}\left (\mathbf{y}_{i}(t) \right ) \right) - x_{i}(t) , \\
		\label{y_dyn} \dot{y}_{ij}(t) =& - \beta \Bigg[ {\sum_{k=1}^{N}a_{ik}\left (y_{ij}(t)-y_{kj}(t)\right )}  \nonumber \\
		& + {a_{ij}\left (y_{ij}(t)-x_{j}(t)\right )} \Bigg],
	\end{align}
	where $\alpha, \beta >0$ are step sizes, $\mathbf{y}_{i} = [y_{i1}, y_{i2}, \dots, y_{iN}] ^T \in \mathbb{R}^{N}$, 
	$y_{ij}$ is player~$i$'s estimate on player~$j$, $y_{ii} = x_{i}$, and the initial actions are chosen as $x_{i}(0) \in \mathcal{X}_{i}$.
	
	The algorithm composed of~\eqref{x_dyn} and~\eqref{y_dyn} needs continuous communication among players. We employ an event-triggered mechanism to reduce the communication times, i.e., a player only broadcasts its state and estimate on all players when certain critical events occur. The update law of~\eqref{y_dyn} becomes
	\begin{equation}\label{y_dyn_et} 
		\begin{aligned}
			\dot{y}_{ij}(t) =& - \beta \Bigg[ \sum_{k=1}^{N} a_{ik}\left (\hat{y}_{ij}(t)-\hat{y}_{kj}(t)\right )  \\
			& + a_{ij}\left (\hat{y}_{ij}(t)-\hat{x}_{j}(t)\right )\Bigg],
		\end{aligned}
	\end{equation}
	where $\hat{y}_{ij}$ is the latest estimate on player~$j$ broadcast by player~$i$, and $\hat{x}_{j}$ is the latest state broadcast by player~$j$. Suppose the triggering time instants of player~$i$ is $\{ t_{1}^{i}, t_{2}^{i}, \dots, t_{k}^{i}, \dots \}$, and then  $\hat{y}_{ij}(t) = y_{ij}(t_{k}^{i}), \hat{x}_{i}\left (t\right ) = x_{i}(t_{k}^{i})$ for $t \in \left[t_{k}^{i}, t_{k+1}^{i} \right)$. 
	
	Our objective is to develop an event-triggered mechanism such that the NE can be asymptotically achieved. Specifically, we aim to design a decision variable:
	\begin{equation*}
		\gamma_{i}(t) = \left\{
		\begin{array}{cl}
			1, & \hat{x}_{i}(t) = x_{i}(t), \hat{y}_{ij}(t) = y_{ij}(t), \\
			0, & \text{otherwise}, \\
		\end{array}
		\right. \forall i \in \mathcal{V},
	\end{equation*}
	so that the average communication rate
	\begin{equation}
		\Gamma(t) = \frac{1}{Nt} \sum_{i=1}^{N} \int_{0}^{t} \gamma_{i}(t) dt
	\end{equation}
	with $\Gamma(0)=0$ can be reduced. For the stochastic event-triggered mechanism, we consider its expected value $\mathbb{E}[\Gamma(t)]$ due to the randomness of $\gamma_{i}(t)$. Moreover, the event-triggered mechanism should not exhibit Zeno behavior which refers to the phenomenon that an infinite number of events occur in a finite time.
		\begin{rem}
			The problem formulation is different \textcolor{blue}{from} the distributed optimization problem solved in~\cite{tsang2020distributed}, where $N$ agents cooperatively minimize a global cost function, $F(x) = \frac{1}{N} \sum_{i=1}^{N} f_{i}(x)$, and the cost function of agent $i$ is \emph{decoupled} with other agents' actions, $x_{j}$, $j \neq i$. Although~\eqref{eq: DG} can be regarded as a set of parallel optimization problems, each player's cost function $f_{i}$ depends on all other players' decisions $\mathbf{x}_{-i}$. However, each player only accesses information about its neighbors. The players need to keep an estimate on other players' strategies, $\mathbf{y}_{i}$, and communicate this information to neighbors to seek the NE. Owing to this more complex information exchange setting, the stochastic event-triggering law and the convergence analysis become more complex compared to~\cite{tsang2020distributed}.
		\end{rem}

	\section{Main Results} \label{sec: results}
	In this section, a stochastic event-triggered algorithm is proposed for the distributed NE seeking. Then we prove that the NE can be sought with exponential convergence rate without Zeno behavior.
	\subsection{Proposed Stochastic Event-Triggering Law}
	The compact form of~\eqref{x_dyn} and~\eqref{y_dyn_et} can be written as
	\begin{align}
		\label{c_x_dyn_et} \dot{\mathbf{x}}(t) &= \mathbb{P}_{\mathcal{X}} \left( \mathbf{x}(t) - \alpha \frac{\partial{f}}{\partial \mathbf{x}}(\mathbf{y} (t)) \right) - \mathbf{x}(t), \\
		\label{c_y_dyn_et} \dot{\mathbf{y}}(t) 
		& = - \beta \left[\left (L \otimes I_{N} + B_{0}\right ) \left(\hat{\mathbf{y}}(t) - \bm{1}_{N} \otimes \hat{\mathbf{x}}(t)\right)\right],
	\end{align}
	where $\mathbf{x} = \left[x_{1}, x_{2}, \dots, x_{N} \right]^T$, 
	$\hat{\mathbf{x}} = \left[\hat{x}_{1}, \hat{x}_{2}, \dots, \hat{x}_{N}\right]^T$,
	$\mathbf{y} = \left[\mathbf{y}_{1}^{T}, \mathbf{y}_{2}^{T}, \dots, \mathbf{y}_{N}^{T}\right]^{T}$, 
	$\hat{\mathbf{y}} = \left[\hat{\mathbf{y}}_{1}^{T}, \hat{\mathbf{y}}_{2}^{T}, \dots, \hat{\mathbf{y}}_{N}^{T}\right]^{T}$,
	$\hat{\mathbf{y}}_{i} = \left[\hat{y}_{i1}, \hat{y}_{i2}, \dots, \hat{y}_{iN}\right]^{T}$, and $\frac{\partial f}{\partial \mathbf{x}}(\mathbf{y}) = \left[\frac{\partial f_{1}}{\partial x_{1}}(\mathbf{y}_{1}), \frac{\partial f_{1}}{\partial x_{2}}(\mathbf{y}_{2}), \dots, \right. \\
	\left. \frac{\partial f_{1}}{\partial x_{N}}(\mathbf{y}_{N}) \right]^{T}$. 
The equality~\eqref{c_y_dyn_et} holds from $(L \otimes I_{N})(\bm{1}_{N} \otimes \hat{\mathbf{x}}(t)) = 0$.
	
	We define the event errors of player $i$ as
	\begin{align}
		\label{e_x} e_{x_{i}}(t) &= \hat{x}_{i}(t) - x_{i}(t), \\
		\label{e_y} \mathbf{e}_{\mathbf{y}_{i}}(t) &= \hat{\mathbf{y}}_{i}(t) - \mathbf{y}_{i}(t),
	\end{align}
	and the consensus error between player $i$'s estimate and $j$'s estimate as
	\begin{equation}
		\label{consensus_e}   \Delta_{ij}(t)=  \hat{\mathbf{y}}_{i}(t) - \hat{\mathbf{y}}_{j}(t).
	\end{equation}	
	We propose a stochastic event trigger:
	\begin{equation} \label{ET_law}
		\gamma_{i}\left (t\right ) = \left\{
		\begin{array}{cl}
			1, & \xi_{i}(t) > \kappa \exp\left (-c_{i} \rho_{i} (t)/ \delta_{i} (t)\right ), \\
			0, & \text{otherwise}, \\
		\end{array}
		\right.
	\end{equation}
	where $\kappa>1$ is a parameter, $\xi_{i}\left (t\right ) \in (a, 1)$ an arbitrary stationary ergodic random process with a constant $a>0$, $c_{i}>0$ a constant, and $\delta_{i}(t)>0$ a decreasing function w.r.t. $t$. Inspired by~\cite{zhang2021distributed} and~\cite{tsang2020distributed}, $\rho_{i}(t)$ and $\delta_{i}(t)$ are defined as
	\begin{align}
		\label{eq: rho}\rho_{i} (t) &= e_{x_i} (t)^{2} + \left \| \mathbf{e}_{\mathbf{y}_{i}} (t) \right \| ^{2} - \sigma_{i}\left \| \sum_{j=1}^{N} a_{ij} \Delta_{ij} (t) \right \| ^{2}, \\
		\dot{\delta}_{i}(t) &= -\eta \delta_{i} (t),
		\label{delta}
	\end{align}
	where $\sigma_{i}>0$ and $\eta>0$.
	According to~\eqref{ET_law} and~\eqref{eq: rho}, we can infer the following condition when no trigger occurs, i.e., $\gamma_{i}(t) = 0$:
	\begin{equation} \label{no_trigger_con}
		\begin{split}
			& e_{x_i} (t )^{2} + \left \| \mathbf{e}_{\mathbf{y}_{i}} (t) \right \| ^{2} - \sigma_{i} \left \| \sum_{j=1}^{N} a_{ij} \Delta_{ij} (t) \right \| ^{2} \\
			\leq & \frac{\delta_{i} (t )}{c_{i}} \left (\ln\kappa - \ln\xi_{i} (t )\right ).
		\end{split}
	\end{equation} 
	
	\begin{rem}
		In the literature, $\rho_{i} (t)$ is usually called a triggering function, depending on event error, consensus error, and network parameters. Different triggering functions lead to different event-triggering laws, yielding different performances. In deterministic event-triggered mechanisms, player~$i$ triggers always whenever $\rho_{i}(t)>0$~(\cite{yi2018dynamic,singh2022sparq, zhang2021distributed}). However, in stochastic event triggers, player~$i$ triggers with a certain probability which increases with $\rho_{i}(t)$. 
			As an illustration, we consider a case where $a=\frac{1}{2}$, and $\xi_{i}(t)$ is a uniformly distributed random process. When $\rho_{i}(t)\leq 0$, there is $\kappa \exp \left( -c_{i}\rho_{i}(t)/ \delta_{i}(t) \right) \geq 1 > \xi_{i}(t)$, which makes it impossible for player $i$ to trigger. This is consistent with the deterministic event-triggering law as $\gamma_{i}(t)=0$ when $\rho_{i}(t) \leq 0$.
			 If $\rho_{i}(t) > 0$, we can infer that $P \left[ \gamma_{i}(t)=1 \right] = \frac{1}{2} \left[ 1- \kappa \exp \left( -c_{i}\rho_{i}(t)/ \delta_{i}(t) \right) \right]$ based on the distribution of $\xi_{i}(t)$, i.e., $P \left[ \gamma_{i}(t)=1 \right]$ is monotonically increasing with the value of $\rho_{i}(t)$.
		When $\xi_{i}(t)$ is a strictly positive constant, the stochastic event trigger becomes deterministic. In this sense,~\eqref{ET_law} can be regarded as a generalized version of the deterministic trigger, further reducing communication burden. It is suitable for networks with tighter communication requirements, thus more practical. 
	\end{rem}
	
	\subsection{Convergence Analysis}
	To simplify notation, the time index $t$ is omitted in the following analysis.
	
	Define the seeking errors of $\mathbf{x}$ and $\mathbf{y}$ as $\bm{\varepsilon}_{\mathbf{x}} = \mathbf{x} - \mathbf{x}^{*}$ and $\bm{\varepsilon}_{\mathbf{y}} = \mathbf{y} - \mathbf{1}_{N} \otimes \mathbf{x}$. Then based on~\eqref{c_x_dyn_et}--\eqref{e_y}, the dynamics of $\bm{\varepsilon}_{\mathbf{x}}$ and $\bm{\varepsilon}_{\mathbf{y}}$ can be written as
	\begin{align*}
		\dot{\bm{\varepsilon}}_{\mathbf{x}}=& \dot{\mathbf{x}} =  \mathbb{P}_{\mathcal{X}} \left( \mathbf{x} - \alpha \frac{\partial{f}}{\partial \mathbf{x}} (\bm{\varepsilon}_{\mathbf{y}} + \mathbf{1}_{N} \otimes \mathbf{x}) \right) - \mathbf{x}, \\
		\dot{\bm{\varepsilon}}_{\mathbf{y}} =&\dot{\mathbf{y}} - \bm{1}_{N} \otimes \dot{\mathbf{x}}\\
		=& \beta \left (L \otimes I_{N} + B_{0}\right ) \left (\bm{1}_{N} \otimes \mathbf{e}_{\mathbf{x}} - \bm{\varepsilon}_{\mathbf{y}} - \mathbf{e}_{\mathbf{y}} \right ) \\ 
		&-\bm{1}_{N} \otimes \Bigg\{ \mathbb{P}_{\mathcal{X}} \left( \mathbf{x} - \alpha \frac{\partial{f}}{\partial \mathbf{x}} (\bm{\varepsilon}_{\mathbf{y}} + \mathbf{1}_{N} \otimes \mathbf{x}) \right)  - \mathbf{x} \Bigg\},
	\end{align*}
where $\mathbf{e}_{\mathbf{x}}=[e_{x_{1}}, e_{x_{2}}, \dots, e_{x_{N}}]^{T}$ and $\mathbf{e}_{\mathbf{y}} = [\mathbf{e}_{\mathbf{y}_{1}}^{T}, \mathbf{e}_{\mathbf{y}_{2}}^{T}, \dots, \\ \mathbf{e}_{\mathbf{y}_{N}}^{T}]^{T}$.
	
	\begin{thm} \label{thm: convergence} 
		For a multi-agent system, suppose that Assumptions~\ref{existence} and~\ref{unique} are satisfied. The distributed algorithm~\eqref{x_dyn} and~\eqref{y_dyn_et} under the stochastic event-triggering law~\eqref{ET_law} exponentially converges to the NE $\mathbf{x}^{*}$ with 
		\begin{align}
			\label{alpha_b} 0< \alpha &< \frac{2\mu \beta \lambda - 8C_{2}C_{3} - 2\mu C_{4}}{8C_{1}C_{2}C_{3} + 4\mu C_{2}C_{3} + \beta C_{1}^{2} \lambda - C_{1}^{2}C_{4}} , \\
			\label{beta_b} \beta &> \frac{4C_{2}C_{3}+\mu C_{4}}{\mu \lambda}, \\
			0<\sigma &\leq \frac{N-1}{2N \left\| L \right\|^{2}},
		\end{align}
		where $\sigma = \max_{i} \{ \sigma_{i} \}$, $\lambda$ is the minimum eigenvalue of $Q$, $C_{1} = \bar{l}\sqrt{N}$, $C_{2} = \bar{l}$, $C_{3} = \sqrt{N}\left\|P \right\|$, $C_{4} = 2\sqrt{2(N-1)}\left\| P(L\otimes I_{N} + B_{0}) \right\|$, and $\bar{l} = \max_{i}\{l_{i}\}$.
	\end{thm}
	\begin{pf}
		Inspired by~\cite{liang2022exponentially}, we consider the Lyapunov candidate
		\begin{equation} \label{Lyapunov}
			V = \phi_{2} V_1 + \phi_{1} V_2 + \zeta\sum_{i = 1}^{N}\delta_{i},
		\end{equation}
		where $V_{1}=  \bm{\varepsilon}_{\mathbf{x}}^{T}\bm{\varepsilon}_{\mathbf{x}}, V_{2} = \bm{\varepsilon}_{\mathbf{y}}^{T}P\bm{\varepsilon}_{\mathbf{y}}$, $\phi_{1} = 2\alpha C_{2}$, $\phi_{2} = 2C_{3}(2+\alpha C_{1})$, $\zeta= \frac{2\phi_{1}C_{5}}{\eta \min_{i}\{c_{i}\}}(\ln \kappa - \ln a)$, and $C_{5} = N\sqrt{\frac{2}{N-1}}\left \| P \left (L \otimes I_{N} + B_{0}\right ) \right \|$.
		
		
		For the time derivative of $V_{1}$, we have
		\begin{align} \label{V1_dot}
			 \dot{V}_{1}(t) 
			= & 2 \bm{\varepsilon}_{\mathbf{x}} (t)^{T} \dot{\bm{\varepsilon}}_{\mathbf{x}} (t) \nonumber \\
			= & -2  \bm{\varepsilon}_{\mathbf{x}}^{T} \Bigg\{
			\mathbf{x} - \mathbb{P}_{\mathcal{X}} \left( \mathbf{x} - \alpha \frac{\partial{f}}{\partial \mathbf{x}}(\mathbf{1}_{N}\otimes \mathbf{x}) \right)  \Bigg\} \nonumber \\
			& -2  \bm{\varepsilon}_{\mathbf{x}}^{T} \Bigg\{
			\mathbf{x} - \mathbb{P}_{\mathcal{X}} \left( \mathbf{x} - \alpha \frac{\partial{f}}{\partial \mathbf{x}}( \bm{\varepsilon}_{\mathbf{y}}  + \mathbf{1}_{N}\otimes \mathbf{x}) \right) \nonumber \\
			&- \bigg[ \mathbf{x} - \mathbb{P}_{\mathcal{X}} \left( \mathbf{x} - \alpha \frac{\partial{f}}{\partial \mathbf{x}}(\mathbf{1}_{N}\otimes \mathbf{x}) \right) \bigg]  \Bigg\}. 
		\end{align}
		For the first term of~\eqref{V1_dot},
		\begin{align}\label{V_11}
			&2  \bm{\varepsilon}_{\mathbf{x}}^{T} \Bigg\{
			\mathbf{x} - \mathbb{P}_{\mathcal{X}} \left( \mathbf{x} - \alpha \frac{\partial{f}}{\partial \mathbf{x}}(\mathbf{1}_{N}\otimes \mathbf{x}) \right)  \Bigg\} \nonumber \\
			= & 2  \bm{\varepsilon}_{\mathbf{x}}^{T} \Bigg\{
			\mathbf{x} - \mathbb{P}_{\mathcal{X}} \left( \mathbf{x} - \alpha \frac{\partial{f}}{\partial \mathbf{x}}(\mathbf{1}_{N}\otimes \mathbf{x}) \right) \nonumber \\
			&- \bigg[ \mathbf{x}^{*} - \mathbb{P}_{\mathcal{X}} \left( \mathbf{x}^{*} - \alpha \frac{\partial{f}}{\partial \mathbf{x}}(\mathbf{1}_{N}\otimes \mathbf{x}^{*}) \right) \bigg] \Bigg\} \nonumber \\
			= & 2 \left\| \bm{\varepsilon}_{\mathbf{x}} \right\|^{2} - 2 \bm{\varepsilon}_{\mathbf{x}}^{T}\bigg[ \mathbb{P}_{\mathcal{X}} \left( \mathbf{x} - \alpha \frac{\partial{f}}{\partial \mathbf{x}}(\mathbf{1}_{N}\otimes \mathbf{x}) \right) \nonumber \\
			&- \mathbb{P}_{\mathcal{X}} \left( \mathbf{x}^{*} - \alpha \frac{\partial{f}}{\partial \mathbf{x}}(\mathbf{1}_{N}\otimes \mathbf{x}^{*}) \right)  \bigg] \nonumber \\
			\geq & 2 \left\| \bm{\varepsilon}_{\mathbf{x}} \right\| \left(  \left\|\bm{\varepsilon}_{\mathbf{x}}\right\| - \left\|  \bm{\varepsilon}_{\mathbf{x}} - \alpha d(\mathbf{x}) \right\| \right),
		\end{align}
		where $d(\mathbf{x}) =  \frac{\partial{f}}{\partial \mathbf{x}}(\mathbf{1}_{N}\otimes \mathbf{x}) - \frac{\partial{f}}{\partial \mathbf{x}}(\mathbf{1}_{N}\otimes \mathbf{x}^{*})$, and the first equation holds from~\eqref{eq: NE_property}. Then
		\begin{align}\label{V_111}
			 \left\|\bm{\varepsilon}_{\mathbf{x}}\right\| - \left\|  \bm{\varepsilon}_{\mathbf{x}} - \alpha d(\mathbf{x}) \right\|\nonumber 
			= & \frac{\left\|\bm{\varepsilon}_{\mathbf{x}}\right\|^{2} - \left\|  \bm{\varepsilon}_{\mathbf{x}} - \alpha d(\mathbf{x}) \right\|^{2}}
			{\left\|\bm{\varepsilon}_{\mathbf{x}}\right\| + \left\|  \bm{\varepsilon}_{\mathbf{x}} - \alpha d(\mathbf{x}) \right\|} \nonumber \\
			\geq & \frac{2 \alpha \bm{\varepsilon}_{\mathbf{x}}^{T} d(\mathbf{x}) - \alpha^{2} \left\| d(\mathbf{x}) \right\|^{2}}
			{\left( 2+ \alpha \sqrt{N}\bar{l} \right)\left\| \bm{\varepsilon}_{\mathbf{x}} \right\|} \nonumber\\
			\geq & \frac{2 \alpha \mu - \alpha^{2} \bar{l}^{2}N}{2+ \alpha \bar{l}\sqrt{N}} \left\| \bm{\varepsilon}_{\mathbf{x}} \right\|.
		\end{align}
		For the last two \textcolor{blue}{terms} of~\eqref{V1_dot}, 
		\begin{align}\label{V_12}
			&-2  \bm{\varepsilon}_{\mathbf{x}}^{T} \Bigg\{
			\mathbf{x} - \mathbb{P}_{\mathcal{X}} \left( \mathbf{x} - \alpha \frac{\partial{f}}{\partial \mathbf{x}}( \bm{\varepsilon}_{\mathbf{y}}  + \mathbf{1}_{N}\otimes \mathbf{x}) \right) \nonumber \\
			&- \bigg[ \mathbf{x} - \mathbb{P}_{\mathcal{X}} \left( \mathbf{x} - \alpha \frac{\partial{f}}{\partial \mathbf{x}}(\mathbf{1}_{N}\otimes \mathbf{x}) \right) \bigg]  \Bigg\} \nonumber \\
			\leq & 2 \alpha \left\| \bm{\varepsilon}_{\mathbf{x}} \right\| \left\| \frac{\partial{f}}{\partial \mathbf{x}}( \bm{\varepsilon}_{\mathbf{y}}  + \mathbf{1}_{N}\otimes \mathbf{x})- \frac{\partial{f}}{\partial \mathbf{x}}(\mathbf{1}_{N}\otimes \mathbf{x}) \right\| \nonumber \\
			\leq & 2 \alpha \bar{l} \left\|  \bm{\varepsilon}_{\mathbf{x}} \right\| \left\|  \bm{\varepsilon}_{\mathbf{y}} \right\|
		\end{align}
		Combining~\eqref{V_11},~\eqref{V_111},~\eqref{V_12} into~\eqref{V1_dot}, we get
		\begin{equation} 
			\dot{V}_{1} \leq - \omega_{1} \left\| \bm{\varepsilon}_{\mathbf{x}} \right\|^{2} + \phi_{1} \left\| \bm{\varepsilon}_{\mathbf{x}} \right\|\left\| \bm{\varepsilon}_{\mathbf{y}} \right\|,
			\label{eq: V_1_dot}
		\end{equation}
		where 
		$\omega_{1} = \frac{2(2 \alpha \mu - \alpha^{2} C_{1}^{2})}{2+\alpha C_{1}}$.

		Moreover, the time derivative of $V_{2}$ is
		\begin{align} 
			& \dot{V}_{2}(t) \nonumber \\
			=& 2  \bm{\varepsilon}_{\mathbf{y}}^{T}P \dot{\bm{\varepsilon}}_{\mathbf{y}} \nonumber \\
			=& -2  \bm{\varepsilon}_{\mathbf{y}}^{T}P \Bigg\{ \mathbf{1}_{N} \otimes   \bigg[ \mathbb{P}_{\mathcal{X}} \left( \mathbf{x} -\alpha \frac{\partial{f}}{\partial \mathbf{x}} (\bm{\varepsilon}_{\mathbf{y}} + \mathbf{1}_{N} \otimes \mathbf{x}) \right) - \mathbf{x} \bigg]  \Bigg\} \nonumber \\
			& +2 \beta \bm{\varepsilon}_{\mathbf{y}}^{T}P \left (L \otimes I_{N} + B_{0}\right )\left (\bm{1}_{N} \otimes \mathbf{e}_{\mathbf{x}}-\bm{\varepsilon}_{\mathbf{y}} -\mathbf{e}_{\mathbf{y}}\right ) \nonumber \\
			=&- 2  \bm{\varepsilon}_{\mathbf{y}}^{T}P\Bigg\{ \mathbf{1}_{N} \otimes \bigg[ \mathbb{P}_{\mathcal{X}} \left( \mathbf{x} - \alpha \frac{\partial{f}}{\partial \mathbf{x}} (\bm{\varepsilon}_{\mathbf{y}} + \mathbf{1}_{N} \otimes \mathbf{x}) \right) - \mathbf{x} \bigg]  \Bigg\} \nonumber \\
			& + 2 \beta \bm{\varepsilon}_{\mathbf{y}}^{T}P \left (L \otimes I_{N} + B_{0}\right )(\bm{1}_{N} \otimes \mathbf{e}_{\mathbf{x}}-\mathbf{e}_{\mathbf{y}}) \nonumber\\
			\label{V2_11}&- \beta \bm{\varepsilon}_{\mathbf{y}}^{T}Q\bm{\varepsilon}_{\mathbf{y}}.
		\end{align}
		\textcolor{blue}{Similar} to~\eqref{V_11}, the first term of~\eqref{V2_11} satisfies
		\begin{align} \label{V2_22}
			& -2  \bm{\varepsilon}_{\mathbf{y}}^{T}P \Bigg\{ \mathbf{1}_{N} \otimes \bigg[ \mathbb{P}_{\mathcal{X}} \left( \mathbf{x} - \alpha \frac{\partial{f}}{\partial \mathbf{x}} (\bm{\varepsilon}_{\mathbf{y}} + \mathbf{1}_{N} \otimes \mathbf{x}) \right) - \mathbf{x} \bigg]  \Bigg\} \nonumber \\
			\leq  & 2 \sqrt{N} \left\| \bm{\varepsilon}_{\mathbf{y}} \right\| \left\|P \right\| \left\| \mathbb{P}_{\mathcal{X}} \left( \mathbf{x} - \alpha \frac{\partial{f}}{\partial \mathbf{x}} (\bm{\varepsilon}_{\mathbf{y}} + \mathbf{1}_{N} \otimes \mathbf{x}) \right) - \mathbf{x} \right. \nonumber \\
			&\left. -\left[  \mathbb{P}_{\mathcal{X}} \left( \mathbf{x}^{*} - \alpha \frac{\partial{f}}{\partial \mathbf{x}} ( \mathbf{1}_{N} \otimes \mathbf{x}^{*}) \right) - \mathbf{x}^{*} \right] \right\| \nonumber \\
			\leq & 2 \sqrt{N}  \left\| P\right\|\left(2+\alpha \bar{l}\sqrt{N}\right)\left\|\bm{\varepsilon}_{\mathbf{x}} \right\| \left\| \bm{\varepsilon}_{\mathbf{y}} \right\| + 2\alpha\bar{l}\sqrt{N}\left\| P\right\| \left\| \bm{\varepsilon}_{\mathbf{y}} \right\|^{2},
		\end{align}
		and the second term of~\eqref{V2_11},
		\begin{align}
			& 2 \bm{\varepsilon}_{\mathbf{y}}^{T}P \left (L \otimes I_{N} + B_{0}\right )\left (\bm{1}_{N} \otimes \mathbf{e}_{\mathbf{x}}\right ) \nonumber \\
			\label{ex} \leq & \frac{1}{\nu} \left \| \bm{\varepsilon}_{\mathbf{y}} \right \| ^{2} + \nu N \left \| P\left (L \otimes I_{N} + B_{0}\right ) \right \| ^{2} \left \| \mathbf{e}_{\mathbf{x}} \right \| ^{2}, 
		\end{align}
		and
		\begin{align}
			& -2 \bm{\varepsilon}_{\mathbf{y}}^{T}P \left (L \otimes I_{N} + B_{0}\right )\mathbf{e}_{\mathbf{y}} \nonumber \\
			\label{ey} \leq & \frac{1}{\nu} \left \| \bm{\varepsilon}_{\mathbf{y}} \right \| ^{2} + \nu \left \| P \left (L \otimes I_{N} + B_{0}\right ) \right \| ^{2} \left \| \mathbf{e}_{\mathbf{y}} \right \| ^{2},
		\end{align}
		for any $\nu >0$ according to Young's inequality. 
		
		According to~\eqref{no_trigger_con},
		\begin{equation*}
			\begin{aligned}
				& \left \| \mathbf{e}_{\mathbf{x}} \right \| ^{2} + \left \| \mathbf{e}_{\mathbf{y}} \right \| ^{2} \\
				= &\sum_{i=1}^{N} e_{x_i}^{2} + \sum_{i=1}^{N}  \left \| \mathbf{e}_{\mathbf{y}_{i}} \right \| ^{2} \\
				\leq & \sum_{i=1}^{N} \frac{\delta_{i}}{c_{i}} \left (\ln \kappa - \ln \xi_{i}\right ) +  \sum_{i=1}^{N} \sigma_{i} \left \| \sum_{j=1}^{N} a_{ij} \Delta_{ij}\left (t\right )\right \| ^{2} \\
				\leq & \sum_{i=1}^{N} \frac{\delta_{i}}{c_{i}} \left (\ln \kappa - \ln \xi_{i}\right ) + 2\sigma \left \| L \right \| ^{2} \left ( \left \| \mathbf{e}_{\mathbf{y}} \right \|^{2} + \left \| \bm{\varepsilon}_{\mathbf{y}} \right \|^{2} \right ).
			\end{aligned}
		\end{equation*}
		Then, we have
		\begin{align*}
			&\left \| \mathbf{e}_{\mathbf{x}} \right \| ^{2} + \left (1-2\sigma \left \| L \right \| ^{2}\right )\left \| \mathbf{e}_{\mathbf{y}} \right \| ^{2} \\
			\leq & \sum_{i=1}^{N} \frac{\delta_{i}}{c_{i}} \left (\ln \kappa - \ln \xi_{i}\right ) + 2\sigma \left \| L \right \| ^{2}  \left \| \bm{\varepsilon}_{\mathbf{y}} \right \|^{2}.
		\end{align*}
		If $\sigma \leq \frac{N-1} {2N \left \| L \right \| ^{2}}$, the combination of the second terms of~\eqref{ex} and~\eqref{ey} satisfies
		\begin{align} \label{V2_12}
			& \nu N \left \| P \left (L \otimes I_{N} + B_{0}\right ) \right \| ^{2} ( \left \| \mathbf{e}_{\mathbf{x}} \right \| ^{2} + \frac{1}{N} \left \| \mathbf{e}_{\mathbf{y}} \right \| ^{2}) \nonumber \\
			\leq & \nu N \left \| P \left (L \otimes I_{N} + B_{0}\right ) \right \| ^{2} \left[\left \| \mathbf{e}_{\mathbf{x}} \right \| ^{2} + \left (1-2\sigma \left \| L \right \| ^{2}\right ) \left \| \mathbf{e}_{\mathbf{y}} \right \| ^{2}\right] \nonumber \\
			\leq & \nu N \left \| P \left (L \otimes I_{N} + B_{0}\right ) \right \| ^{2} \Bigg[\sum_{i=1}^{N}\frac{\delta_{i}}{c_{i}} \left (\ln \kappa - \ln \xi_{i}\right )  \nonumber \\
			&  + 2\sigma \left \| L \right \| ^{2} \left \| \bm{\varepsilon}_{\mathbf{y}} \right \|^{2} \Bigg].
		\end{align}
		Letting $\nu = \frac{\sqrt{2}}{\sqrt{N-1} \left\| P (L \otimes I_{N} + B_{0}) \right\|}$, and combining~\eqref{V2_22}--\eqref{V2_12} into~\eqref{V2_11}, one obtains
		\begin{align}
			\dot{V}_{2} \leq & - \omega_{2} \left\| \bm{\varepsilon}_{\mathbf{y}} \right\|^{2} + \phi_{2} \left\|\bm{\varepsilon}_{\mathbf{x}} \right\| \left\| \bm{\varepsilon}_{\mathbf{y}} \right\| \nonumber \\
			& + C_{5} \Bigg[\sum_{i=1}^{N}\frac{\delta_{i} (t )}{c_{i}} \left (\ln \kappa - \ln \xi_{i}\left (t\right )\right ) \Bigg],
			\label{eq: V_2_dot}
		\end{align}
		where 
		$\omega_{2} = \beta \lambda - 2 \alpha C_{2}C_{3} - C_{4}$.
		
		Combining~\eqref{eq: V_1_dot} and~\eqref{eq: V_2_dot}, we have
		\begin{align}
			& \phi_{2} \dot{V}_{1} + \phi_{1} \dot{V}_{2} \nonumber \\
			\leq & -\phi_{2} \omega_{1} \left\|\bm{\varepsilon}_{\mathbf{x}} \right\|^{2} + 2\phi_{1} \phi_{2} \left\|\bm{\varepsilon}_{\mathbf{x}} \right\|\left\|\bm{\varepsilon}_{\mathbf{y}} \right\| - \phi_{1}\omega_{2} \left\|\bm{\varepsilon}_{\mathbf{y}} \right\|^{2} \nonumber \\
			& + \phi_{1} C_{5} \Bigg[\sum_{i=1}^{N}\frac{\delta_{i} (t )}{c_{i}} \left (\ln \kappa - \ln \xi_{i}\left (t\right )\right ) \Bigg] \nonumber \\
			= & - \Theta^{*} (\phi_{2} \left\|\bm{\varepsilon}_{\mathbf{x}} \right\|^{2} + \phi_{1} \left\|\bm{\varepsilon}_{\mathbf{y}} \right\|^{2}) - \left(\omega_{1} - \Theta^{*} \right) \phi_{2} \left\|\bm{\varepsilon}_{\mathbf{x}} \right\|^{2} \nonumber \\
			&- \left(\omega_{2} - \Theta^{*} \right) \phi_{1} \left\|\bm{\varepsilon}_{\mathbf{y}} \right\|^{2} + 2\phi_{1} \phi_{2} \left\|\bm{\varepsilon}_{\mathbf{x}} \right\|  \left\|\bm{\varepsilon}_{\mathbf{y}} \right\| \nonumber \\
			& + \phi_{1} C_{5} \Bigg[\sum_{i=1}^{N}\frac{\delta_{i} (t )}{c_{i}} \left (\ln \kappa - \ln \xi_{i}\left (t\right )\right ) \Bigg] \nonumber \\
			\leq & - \Theta^{*} \left(\phi_{2} {V}_{1} + \frac{\phi_{1}}{\lambda_{M}(P)} {V}_{2} \right) \nonumber \\
			& + \phi_{1} C_{5}\Bigg[\sum_{i=1}^{N}\frac{\delta_{i} (t )}{c_{i}} \left (\ln \kappa - \ln \xi_{i}\left (t\right )\right ) \Bigg],
		\end{align}
		where $\Theta^{*} = \left( \omega_{1} + \omega_{2} - \sqrt{(\omega_{1} - \omega_{2})^{2} + 4\phi_{1} \phi_{2}} \right) / 2$, and $\lambda_{M}(P)$ is the maximum eigenvalue of $P$. It follows from~\eqref{alpha_b} and~\eqref{beta_b} that $\Theta^{*} >0$, and $(\omega_{1} - \Theta^{*})(\omega_{2} - \Theta^{*}) = \phi_{1} \phi_{2}$.
		Thus, 
		\begin{align} \label{eq: exp_convergence}
			\dot{V} (t ) \leq & -\Theta^{*} \left(\phi_{2} {V}_{1} + \frac{\phi_{1}}{\lambda_{M}(P)} {V}_{2} \right) - \eta \zeta\sum_{i = 1}^{N} \delta_{i} (t ) \nonumber \\
			&+ \phi_{1} C_{5}\Bigg[\sum_{i=1}^{N}\frac{\delta_{i} (t )}{c_{i}} \left (\ln \kappa - \ln \xi_{i}\left (t\right )\right ) \Bigg] \nonumber \\
			\leq & -\Theta^{*} \left(\phi_{2} {V}_{1} + \frac{\phi_{1}}{\lambda_{M}(P)} {V}_{2} \right) - \eta \zeta\sum_{i = 1}^{N} \delta_{i} (t ) \nonumber \\
			& + \frac{\phi_{1} C_{5}}{\min_{i}\{c_{i}\}} \left (\ln\kappa - \ln a\right ) \sum_{i=1}^N \delta_i (t ) \nonumber \\
			= & -\Theta^{*} (\phi_{2} {V}_{1} + \frac{\phi_{1}}{\lambda_{M}(P)} {V}_{2}) - \frac{\eta \zeta}{2} \sum_{i=1}^N \delta_i (t ) \nonumber \\
			\leq & -k_{v} V (t),	
		\end{align}
		where $k_{v} = \min\left\{\Theta^{*}, \frac{\Theta^{*}}{\lambda_{M}(P)}, \frac{\eta}{2} \right\}$.
		By LaSalle's invariance principle~(\cite{book:143832}), we conclude that $\bm{\varepsilon}_{\mathbf{x}}$ and $\bm{\varepsilon}_{\mathbf{y}}$ converge to zero exponentially.
	\end{pf}
	
	\begin{rem}
		From~\eqref{eq: exp_convergence}, we can see that $k_{v}>0$ is the actual lower bound on the convergence rate, and it depends on $\Theta^{*}$, $\lambda_{M}(P)$, and $\eta$. The variable $\delta_{i}(t)$ plays an important role in the convergence analysis. If $\xi_{i}(t)$ is a strictly positive constant, the above analysis will still be applicable since the stochastic event trigger is an extension of the corresponding deterministic event trigger.
	\end{rem}

\begin{rem}
	We prove that the proposed algorithm achieves the exact NE with an exponential convergence rate, while\textcolor{blue}{~\cite{yu2022distributed} and~\cite{xu2022hybrid} showed that their event-triggered algorithms converge to a neighborhood of the NE,} and~\cite{tsang2020distributed} guaranteed that its optimization algorithm converges to the proximity of the optimal point with arbitrary accuracy. From this perspective, our algorithm and analysis provide a better convergence guarantee. 
\end{rem}

	\subsection{Analysis on Zeno Behavior}
	\begin{thm} \label{thm: Zeno}
		For a multi-agent system, the distributed NE algorithm composed of~\eqref{x_dyn} and~\eqref{y_dyn_et} under the stochastic event-triggering law~\eqref{ET_law} does not exhibit Zeno behavior.
	\end{thm}
	
	\begin{pf}
		Inspired from~\cite{yi2018dynamic}, we demonstrate Zeno behavior exclusion by contradiction. Suppose Zeno behavior exists. Then there must exist a player~$i$ such that for some finite $T>0$,
		\begin{equation*}
			\lim_{k \to \infty} t_{k}^{i} = T, \quad \forall i \in \mathcal{V}.
		\end{equation*}
		In Theorem~\ref{thm: convergence}, we have proven the convergence of the algorithm, which indicates that there exists a constant $Z_{0}>0$, such that 
		\begin{equation} \label{deriv_bound} 
			|\dot{x}_{i}\left (t\right ) | \leq Z_{0}, \left \| \dot{\mathbf{y}}_{i}\left (t\right ) \right \| \leq Z_{0}, \quad \forall i \in \mathcal{V}.
		\end{equation}
		Let 
		$h_{0} = \frac{\sqrt{\ln \kappa}}{2\sqrt{2}Z_{0}} e^{-\frac{1}{2}\eta T }>0$.
		Then based on the property of limit, there exists $k_{0}>0$ such that
		\begin{equation*}
			t_{k}^{i} \in [T-h_{0}, T],\quad \forall k \geq k_{0}.
		\end{equation*} 
		Under the stochastic event-triggering law~\eqref{ET_law}, we have
		\begin{align*}
			&\quad e_{x_i}\left (t\right )^{2} +
			\left \| \mathbf{e}_{\mathbf{y}_{i}}\left (t\right ) \right \| ^{2} 
			\\ &\leq \sigma_{i} \left \| \sum_{j=1}^{N} a_{ij} \Delta_{ij}\left (t\right ) \right \| ^{2} 
			+ \delta(t) \left (\ln\kappa - \ln\xi_{i} (t )\right ),
		\end{align*}
		for $t \in \left[ t_{k_{0}}^{i}, t_{k_{0}+1}^{i} \right)$. In order for the player to trigger at $t_{k_{0}+1}^{i}$, a necessary condition is that
		\begin{equation} \label{upb}
			\begin{aligned}
				&e_{x_i}\left (t_{k_{0}+1}^{i-}\right )^{2} + \left \| \mathbf{e}_{\mathbf{y}_{i}}\left (t_{k_{0}+1}^{i-}\right ) \right \| ^{2}  \\
				>& \sigma_{i} \left \| \sum_{j=1}^{N} a_{ij} \Delta_{ij}\left (t_{k_{0}+1}^{i-}\right ) \right \| ^{2}  \\
				& + \delta\left (t_{k_{0}+1}^{i-}\right ) \left (\ln\kappa - \ln\xi_{i}\left (t_{k_{0}+1}^{i-}\right )\right )\\
				>& e^{\eta t_{k_{0}+1}^{i-}} \ln \kappa,
			\end{aligned}
		\end{equation}
		where $t_{k_{0}+1}^{i-}$ is the time instant right before $t_{k_{0}+1}^{i}$.
		According to~\eqref{deriv_bound}, we can derive that
		\begin{align} \label{lob}
			&\quad e_{x_i} \left( t_{k_{0}+1}^{i-} \right)^{2} + \left\| \mathbf{e}_{\mathbf{y}_{i}}\left( t_{k_{0}+1}^{i-} \right) \right\| ^{2} \nonumber  \\
			& =  \left[ x\left( t_{k_{0}}^{i} \right) - x\left( t_{k_{0}+1}^{i-} \right) \right]^{2} + \left\| \mathbf{y} \left( t_{k_{0}}^{i} \right) - \mathbf{y} \left( t_{k_{0}+1}^{i-}\right) \right\|^{2} \nonumber \\
			& = \left( \int_{t_{k_{0}}^{i}}^{t_{k_{0}+1}^{i-}} \dot{x}_{i}\left( t \right) dt \right)^{2} + \left \| \int_{t_{k_{0}}^{i}}^{t_{k_{0}+1}^{i-}} \dot{\mathbf{y}}_{i}\left (t\right ) dt\right \|^{2} \nonumber \\
			& \leq  2\left[Z_{0}\left (t_{k_{0}+1}^{i-} - t_{k_{0}}^{i}\right ) \right]^{2}.
		\end{align}
		
		Combining~\eqref{upb} and~\eqref{lob}, there is
		\begin{equation}
			\sqrt{2}Z_{0}\left (t_{k+1}^{i-} - t_{k}^{i}\right ) \geq \sqrt{\ln \kappa}\, e^{-\frac{1}{2}\eta t_{k_{0}+1}^{i-}} \geq \sqrt{\ln\kappa}\,   e^{-\frac{1}{2} \eta T}.
		\end{equation}
		Then we can infer that
		\begin{equation}
			t_{k_{0}+1}^{i} - t_{k_{0}}^{i} \geq t_{k_{0}+1}^{i-} - t_{k_{0}}^{i} \geq 2h_{0},
		\end{equation}
		which contradicts $t_{k_{0}+1}^{i} \in [T-h_{0}, T]$. Thus, Zeno behavior does not exist. 
	\end{pf}
	\begin{rem}
		Excluding Zeno behavior validates the well-posedness of the proposed stochastic event-triggered algorithm~\eqref{ET_law}. 
	\end{rem}

	\section{Numerical Simulations} \label{sec: simulation}
	We consider the spectrum access game in energy harvesting BSNs introduced in Section~\ref{sec: introduction}. 
	Based on the formulation in~\cite{niyato2007game}, the spectrum access problem can be formulated as an oligopoly market where $N=5$ BSNs compete with each other in terms of leasing the spectrum size $x_{i} \in [0,16]$ supplied by the primary base station to minimize their own cost, and the cost of the spectrum is determined by a pricing function $p_{i}(\mathbf{x}) = m_{i}^{c} + q_{i}\left( \sum_{j=1}^{N} x_{j} \right)^{\tau}$, where $m_{i}^{c}, q_{i} \geq 0$, for $i \in \{1, 2, \dots, 5\}$, and $\tau \geq 1$. With the allocated spectrum, the BSN can improve the transmission performance using the adaptive modulation, and thus it receives the revenue $r_{i}$ per unit of the achievable transmission rate. If each BSN utilizes uncoded quadrature amplitude modulation with a square constellation, the spectral efficiency of the transmission for BSN $i$ is calculated as:
	\begin{equation}
		u_{i} = \log_{2}\left( 1+ \frac{1.5s_{i}}{\ln\left( \frac{0.2}{\textbf{BER}_{i}^{\text{tar}}} \right)}\right),
	\end{equation}
	where $s_{i}$ is the received signal-to-noise ratio (SNR), indicating the quality of the signal received by the health center, and $\textbf{BER}_{i}^{\text{tar}}$ is the target bit error rate level in the single-input single-output Gaussian noise channel. Then, we can obtain the revenue of the BSN $i$ from $r_{i}u_{i}x_{i}$. Hence, the cost of the BSN $i$ can be obtained as:
	\begin{equation}
		f_{i}(\mathbf{x}) = x_{i}p_{i}(\mathbf{x}) - r_{i}u_{i}x_{i}.
	\end{equation}
	
\begin{figure}[htbp]  
		\centering
		\includegraphics[width=0.3\linewidth]{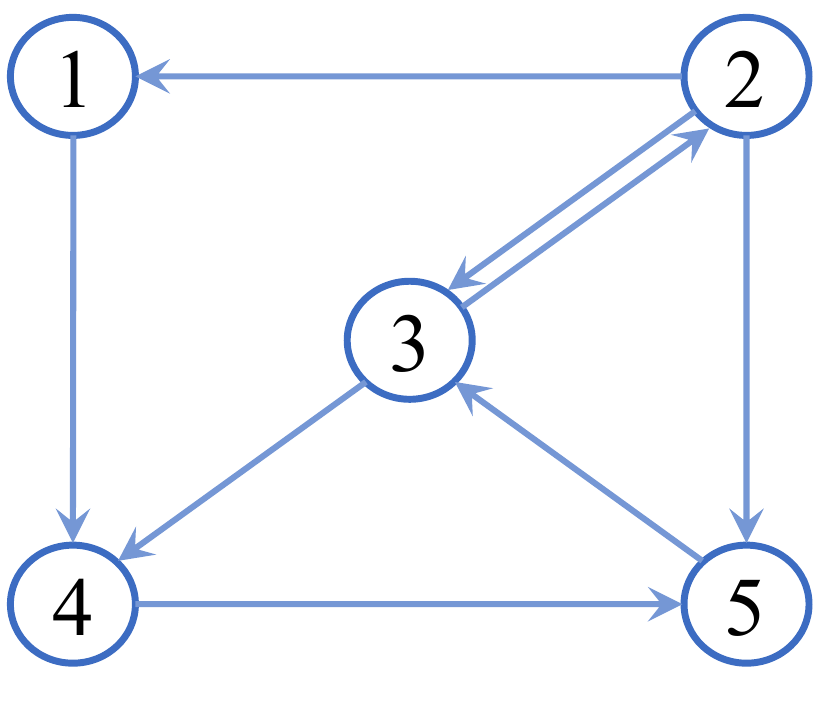}
		\caption{Communication topology for BSNs.} 
		\label{fig: network}
	\end{figure}

	In this simulation, we let BSNs communicate with each other via the directed graph shown in Fig.~\ref{fig: network}.
	We set $a_{ij} = 1$ if $a_{ij}>0$, $\tau = 1$, $\textbf{BER}_{i}^{\text{tar}} = 10^{-4}$, $r_{i} = 20$ for $i = \{1, 2, \dots, 5\}$, $m_{1}^{c} = 5.7$, $m_{2}^{c} = 10.7$, $m_{3}^{c} = 10.3$, $m_{4}^{c} = 9.7$, $m_{5}^{c} = 15$, $q_{1} = 1.1$, $q_{2} = 1.2$, $q_{3} = 1.3$, $q_{4} = 1.4$, $q_{5} = 1.5$, and $s_{1} = 12 \, \si{\decibel}$, $s_{2} = 14 \, \si{\decibel}$, $s_{3} = 15 \, \si{\decibel}$, $s_{4} = 16 \, \si{\decibel}$, $s_{5} = 18 \, \si{\decibel}$. By centralized calculation, we can obtain that the NE for this system is $\mathbf{x}^{*}=[2.000, 3.987, 6.011, 8.018, 9.990 ]^{T}$. The initial actions are \textcolor{blue}{$\mathbf{x}(0) = [14, 12, 10, 4, 2]^{T}$,} and the initial estimates are \textcolor{blue}{$\mathbf{y}_{1}(0) = [0, 1.5, 2.5, 3.5, 4.5]^{T}$, $\mathbf{y}_{2}(0) = [2.5, 3.5, 4.5, 5.5, 6.5]^{T}$, $\mathbf{y}_{3}(0) = [4.5, 5.5, 6.5, 7.5, 8.5]^{T}$, $\mathbf{y}_{4}(0) = [6.5, 7.5, 8.5, 9.5, 10.5]^{T}$, and $\mathbf{y}_{5}(0) = [8.5, 9.5, \\ 10.5, 11.5, 12.5]^{T}$.}
	\textcolor{blue}{We set step sizes as $\alpha = 0.14$, and $\beta = 1.5$}.
	For the stochastic event-triggering law~\eqref{ET_law}, we choose $\eta=10$, $\kappa=1.075$, $\sigma_{i} = 0.8/d_i^{in}$, $c_{i}=1$, and $a=0.05$.
	
\begin{figure}[htbp] \color{blue}
		\centering
		\includegraphics[width=1\linewidth]{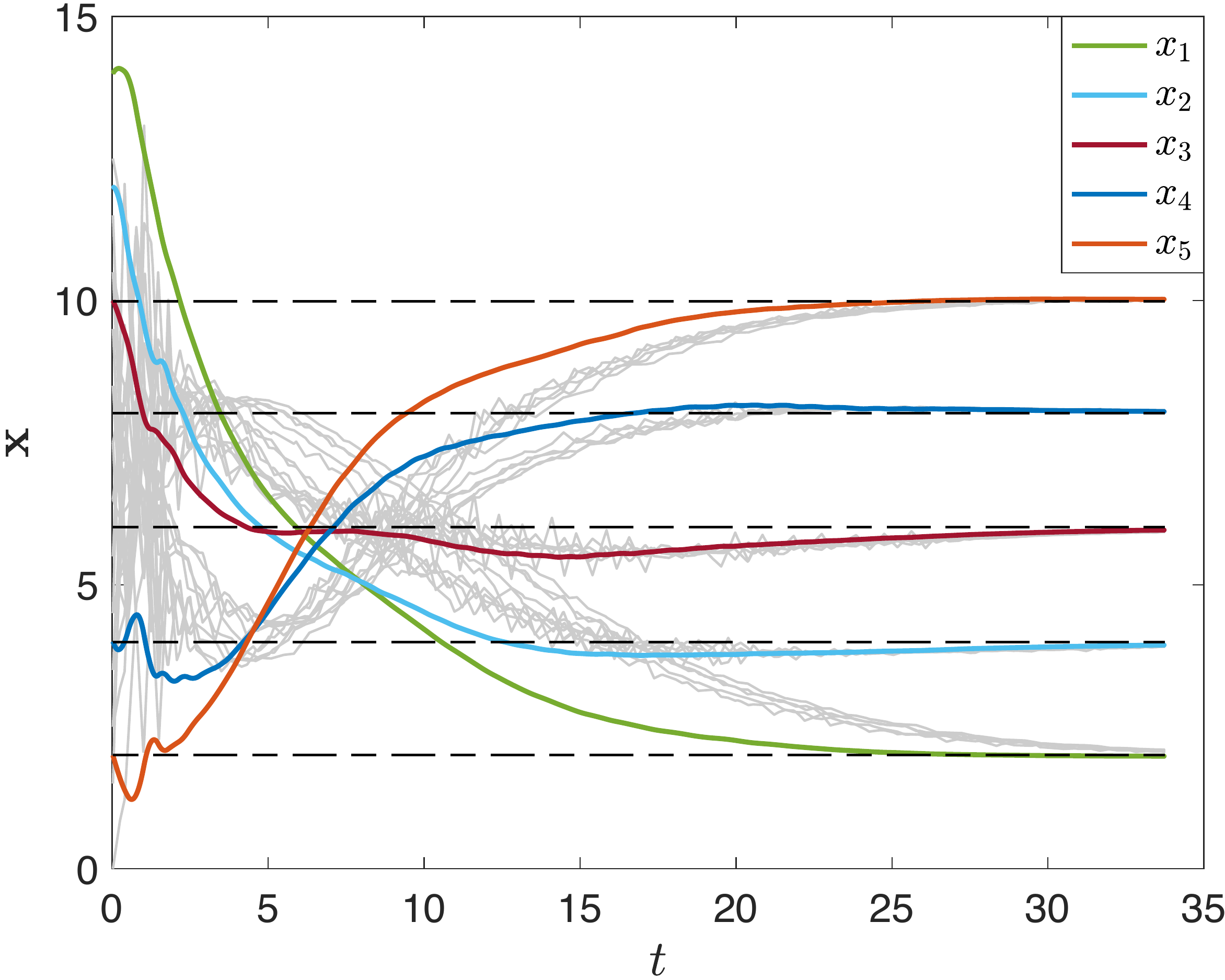}
		\caption{The evolution of BSNs' actions and estimates.} 
		\label{fig: action}
	\end{figure}

\begin{figure}[htbp] \color{blue}
		\centering
		\includegraphics[width=1\linewidth]{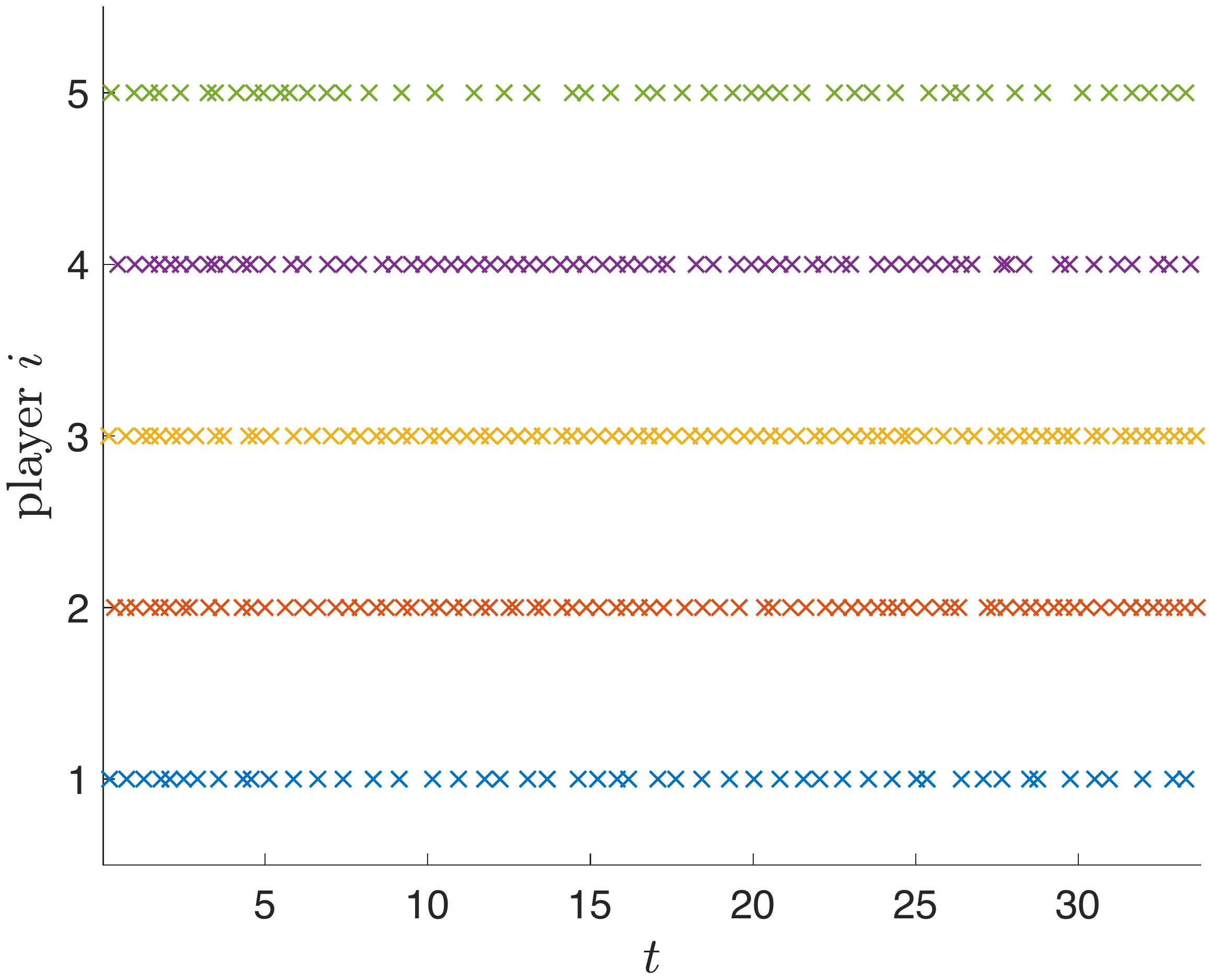}
		\caption{The triggering times for each BSN.} 
		\label{fig: trigger}
	\end{figure}

	\textcolor{blue}{The action and estimates evolutions are illustrated in Fig.~\ref{fig: action}, where the estimates and $\mathbf{x}^{*}$ are shown as blue lines and black dashed lines, respectively,} indicating that all players' actions and estimates converge to the NE as expected. 
	Fig.~\ref{fig: trigger} presents the triggering times for each BSN, illustrating that continuous communication is avoided under~\eqref{ET_law}.
	
	To instantiate the advantage of the proposed stochastic event-triggering law, we compare it with a static one used in~\cite{yu2022distributed}, as well as a dynamic one proposed in~\cite{zhang2021distributed}.
	\begin{figure}[htbp] \color{blue}
		\centering
		\includegraphics[width=1\linewidth]{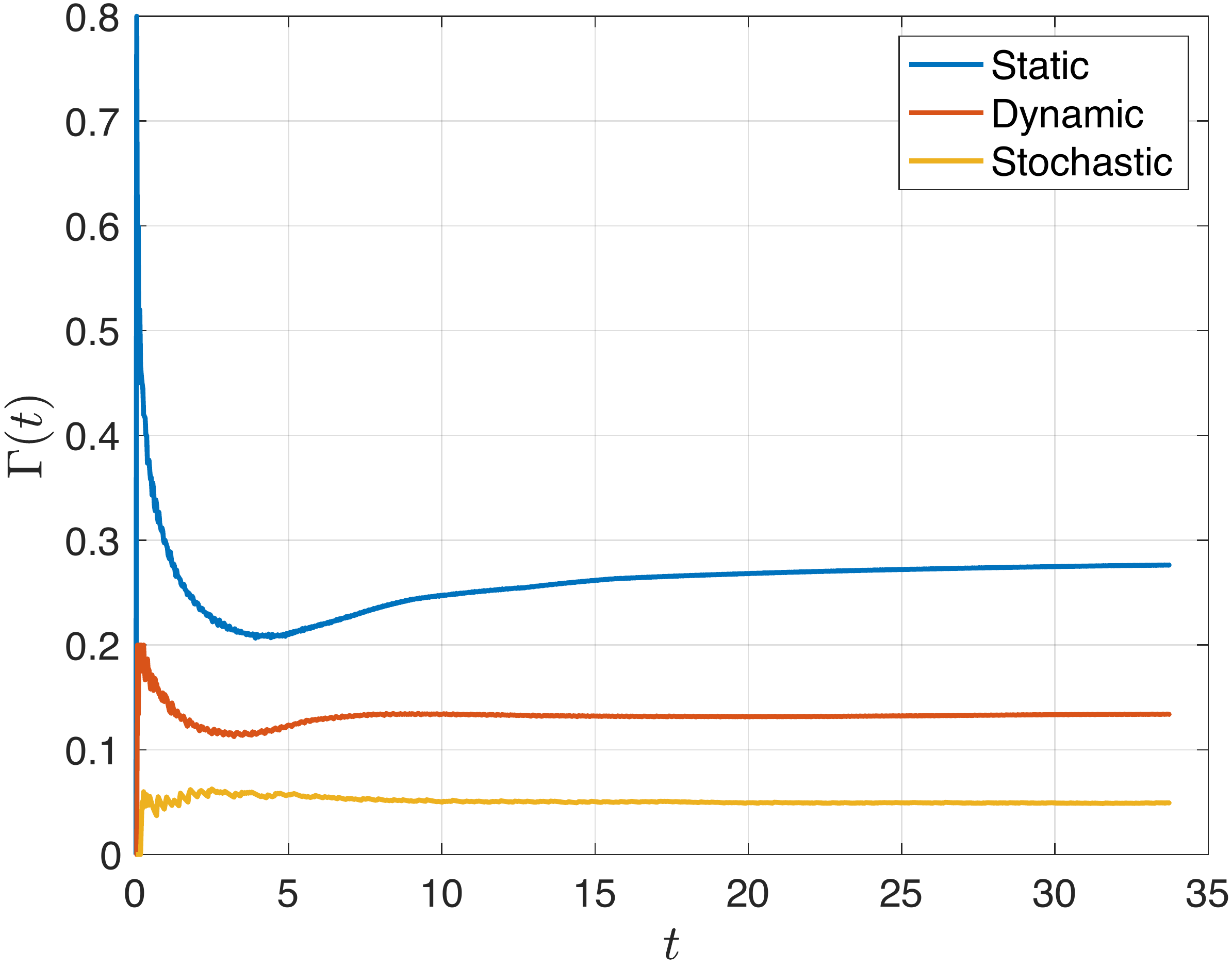}
		\caption{Average communication rates for the static, dynamic and stochastic event-triggering laws.} 
		\label{fig: rate}
	\end{figure}
	Due to the randomness of~\eqref{ET_law}, we run the simulation $100$ times and obtain the empirical mean. 
	The comparison of average communication rates 
	is shown in Fig.~\ref{fig: rate}. 
	We observe that~\eqref{ET_law} achieves the lowest $\Gamma(t)$. Besides, for the stochastic event-triggered mechanism, the peak of the average communication rate is much lower than those of others, implying that the bandwidth can be significantly reduced under~\eqref{ET_law}. 
	Moreover, to observe the triggering times and communication intervals of the three event-triggering laws more intuitively, some involved metrics for all players are summarized in Table~\ref{tab: interval}, indicating that triggering times of~\eqref{ET_law} are mostly fewer and communication intervals are mostly larger than those of other laws.
	\textcolor{blue}{
	\begin{table}[htbp] \color{blue}
		\centering
		\caption{\label{tab: interval}Triggering counts and statistics of communication intervals for players.}   
		\begin{tabular}{p{0.042\textwidth}<{\centering} p{0.065\textwidth}<{\centering} p{0.04\textwidth}<{\centering} p{0.04\textwidth}<{\centering} p{0.04\textwidth}<{\centering} p{0.04\textwidth}<{\centering} p{0.04\textwidth}<{\centering}}    
			\toprule 
			&Player & $1$ & $2$ & $3$ & $4$ & $5$\\    
			\midrule
			\multirow{3}{*}{\shortstack{Trigger \\ count}}
			&Static& $430$ & $420$ & $392$ & $215$ & $407$ \\   
			&Dynamic& $220$ & $185$ & $172$ & $92$ & $234$\\ 
			&Stochastic& $\bf{50}$ & $\bf{81}$ & $\bf{82}$ & $\bf{70}$ & $\bf{49}$ \\   
			\midrule   
			\multirow{3}{*}{\shortstack{Max \\ interval}}
			&Static& $0.100$ & $0.150$ & $0.175$ & $0.225$ & $0.150$ \\   
			&Dynamic& $0.325$ & $0.200$ & $0.275$ & $0.550$ & $0.225$\\ 
			&Stochastic& $\bf{1.050}$ & $\bf{0.875}$ & $\bf{0.775}$ & $\bf{1.125}$ & $\bf{1.250}$ \\ 
			\midrule   
			\multirow{3}{*}{\shortstack{Mean \\ interval}}
			&Static& $0.078$ & $0.080$ & $0.086$ & $0.157$ & $0.083$ \\   
			&Dynamic& $0.154$ & $0.182$ & $0.197$ & $0.367$ & $0.144$\\ 
			&Stochastic& $\bf{0.675}$ & $\bf{0.416}$ & $\bf{0.413}$ & $\bf{0.478}$ & $\bf{0.688}$ \\  
			\midrule 
			\multirow{3}{*}{\shortstack{Min \\ interval}}
			&Static& $0.050$ & $0.050$ & $0.050$ & $0.075$ & $0.050$ \\   
			&Dynamic& $0.125$ & $0.125$ & $0.125$ & $0.200$ & $0.075$\\ 
			&Stochastic& $\bf{0.250}$ & $\bf{0.200}$ & $\bf{0.200}$ & $\bf{0.200}$ & $\bf{0.225}$ \\  
			\midrule 
		\end{tabular}  
	\end{table} 
}

	Intuitively, the stochastic event-triggered algorithm can reduce the communication cost by relaxing the triggering conditions. It is difficult to mathematically characterize the communication rates under different event-triggering laws since this is equivalent to calculate the frequency of $\rho_{i}>0$ for deterministic event triggers or $\xi_{i}(t)>\kappa \exp(-c_{i}\rho_{i}(t)/ \delta_{i}(t))$ for stochastic ones. Therefore, in the literature, most works on distributed algorithms with event-triggered mechanisms only provide convergence analysis without theoretical estimates on the communication rate~(\cite{nowzari2019event, yi2018dynamic,tsang2019zeno, tsang2020distributed, cao2020decentralized, zhang2021distributed, qian2021design, zhao2021distributed, xia2022distributed}). Usually, numerical simulations are provided to illustrate the reduction of communication cost through the proposed event trigger, as we do in our work.		
		Although some works offer the lower bounds of the minimum inter-event time by proving the exclusion of Zeno behavior, these bounds are too loose to be compared~(\cite{nowzari2019event, tsang2020distributed, qian2021design, zhao2021distributed}). 
It is a challenging future work to quantify the communication rate accurately.
	
\begin{figure}[htbp] \color{blue}  
		\centering
		\includegraphics[width=1\linewidth]{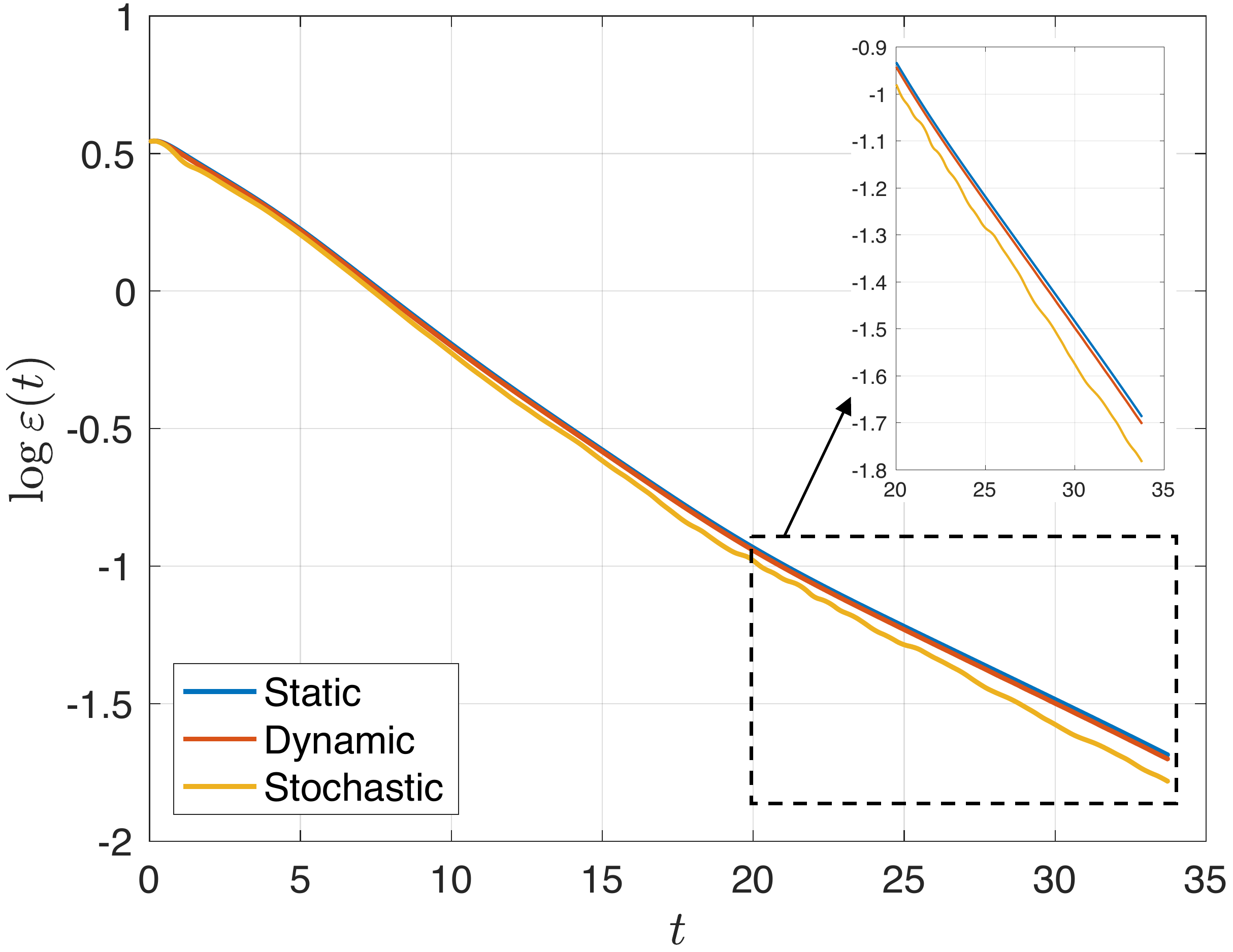}
		\caption{Convergence performance for the static, dynamic and stochastic event-triggering laws in semi-log scale.} 
		\label{fig: error}
	\end{figure}

	
	\textcolor{blue}{Fig.~\ref{fig: error} shows that~\eqref{ET_law} preserves comparable convergence performances even with a slightly faster convergence rate under much lower communication cost.
	In other words,~\eqref{ET_law} can better balance communication efficiency and convergence performance.}
	

	\section{Conclusion and Future Work} \label{sec: conclusions}
	In this paper, we proposed a novel stochastic event-triggered algorithm for the distributed constrained NE seeking problem to improve communication efficiency. In particular, a player transmits its message with a probability increased with the value of the triggering function. We proved the exponential convergence to the exact NE and the non-existence of Zeno behavior. Numerical examples illustrate the significance of our proposed algorithm in practical applications, including much lower communication rates and a slightly faster convergence rate. 
	
	Potential future work includes the rigorous analysis of a tradeoff between the communication rate and the convergence rate, as well as a systematic design of the parameters in the algorithm. 
	
	\bibliographystyle{automatica}        
	\bibliography{autosam}           
	
\end{document}